\documentclass[fleqn]{llncs}

\usepackage{latexsym,amssymb,amsmath,ifthen,alltt,array}
\usepackage[all]{xy}

\renewcommand{\AA}{\mathcal{A}}
\newcommand{\BB}{\mathcal{B}}
\newcommand{\CC}{\mathcal{C}}
\newcommand{\DD}{\mathcal{D}}
\newcommand{\FF}{\mathcal{F}}
\newcommand{\GG}{\mathcal{G}}
\newcommand{\OO}{\mathcal{O}}
\newcommand{\RR}{\mathcal{R}}
\renewcommand{\SS}{\mathcal{S}}
\newcommand{\TT}{\mathcal{T}}
\newcommand{\VV}{\mathcal{V}}
\newcommand{\Var}{\VV\mathsf{ar}}
\newcommand{\Dom}{\DD\mathsf{om}}
\newcommand{\RRs}{\RR_\mathrm{s}}
\newcommand{\RRnv}{\RR_\mathrm{nv}}
\newcommand{\RRg}{\RR_\mathrm{g}}
\newcommand{\REDEX}{\mathsf{REDEX}}
\newcommand{\RS}{\mathsf{RS}}
\newcommand{\NF}{\mathsf{NF}}
\newcommand{\CBNNF}{\mathsf{CBN\mbox{-}NF}}
\newcommand{\CBNNFs}{\CBNNF_\mathrm{s}}
\newcommand{\CBNNFnv}{\CBNNF_\mathrm{nv}}
\newcommand{\CBNNFg}{\CBNNF_\mathrm{g}}
\newcommand{\CBNRS}{\mathsf{CBN\mbox{-}RS}}
\newcommand{\state}[1]{\langle #1 \rangle}
\newcommand{\new}[1]{\langle #1 \rangle'}
\newcommand{\seq}[2][n]{{#2_1},\dots,{#2_{#1}}}

\begin{document}

\bibliographystyle{plain}

\title{On the Complexity of Deciding Call-by-Need
}

\titlerunning{On the Complexity of Deciding Call-by-Need}

\author{
Ir\`ene Durand\inst{1}
\and
Aart Middeldorp\inst{2}
}

\authorrunning{Ir\`ene Durand, Aart Middeldorp}

\tocauthor{
Ir\`ene Durand (Universit\'e de Bordeaux I),
Aart Middeldorp (University of Tsukuba)
}

\institute{
Universit\'e de Bordeaux I \\
33405 Talence, France \\
\email{idurand@labri.u-bordeaux.fr} \\[1ex]
\and
Institute of Information Sciences and Electronics\\
University of Tsukuba, Tsukuba 305-8573, Japan\\
\email{ami@is.tsukuba.ac.jp}
}

\date{\today}
\maketitle

\begin{abstract}

In a recent paper we introduced a new framework for the study of call
by need computations to normal form and root-stable form in term
rewriting. Using elementary tree automata techniques and ground tree
transducers we obtained simple decidability proofs for classes of
rewrite systems that are much larger than earlier classes defined using
the complicated sequentiality concept.
In this paper we show that we can do without ground tree transducers in
order to arrive at decidability proofs that are phrased in direct tree
automata constructions. This allows us to derive better complexity
bounds.
\end{abstract}

\section{Introduction}

The seminal work of Huet and L\'evy~\cite{HL79} on optimal normalizing
reduction strategies for orthogonal rewrite systems marks the beginning
of the quest for decidable subclasses of (orthogonal) rewrite systems
that admit a computable call by need strategy for deriving normal
forms. Call by need means that the strategy may only contract
\emph{needed} redexes, i.e., redexes that are contracted in every
normalizing rewrite sequence. Huet and L\'evy showed that for the
class of orthogonal rewrite systems every term not in normal form
contains a needed redex and repeated contraction of needed redexes
results in a normal form if the term under consideration has a normal
form. However, neededness is in general undecidable. In order to obtain
a decidable approximation to neededness Huet and L\'evy introduced in
the second part of \cite{HL79} the subclass of
\emph{strongly sequential} systems. In a strongly sequential system
at least one of the needed redexes in every reducible term can be 
effectively computed. Moreover, Huet and L\'evy showed that strong
sequentiality is a decidable property of orthogonal rewrite systems.

Strong sequentiality is determined by the left-hand sides of a rewrite
system. By incorporating information of the right-hand sides,
Oyamaguchi~\cite{O93} showed that the class of strongly sequential
systems can be enlarged without losing its good properties. The
resulting class of NV-sequential systems was slightly extended by
Nagaya \textsl{et al.}~\cite{NST95}. Comon~\cite{C95} connected
sequentiality notions with tree automata techniques, resulting in much
shorter decidability proofs for larger classes of rewrite systems.
Jacquemard~\cite{J96} built upon the work of Comon. His class of
growing-sequential rewrite systems extends all previously defined
classes while still being decidable.

In a previous paper (Durand and Middeldorp~\cite{DM97}) we presented a
new framework for decidable call by need. This framework, which we
briefly recall in the next section, is simpler because complicated
notions like sequentiality and index are avoided and \emph{hence} more
powerful. Moreover, we showed how to eliminate the difficult connection
between tree automata and definability in weak second-order monadic
logic in \cite{C95,J96} by assigning a greater role to the concept of
ground tree transducer (GTT, \cite{DHLT90}). In this paper we show that
we can do without GTTs as well.

Not much is known about the complexity of the problem of deciding
membership in one of the classes that guarantees a computable call by
need strategy to normal form. Comon~\cite{C95} showed that strong
sequentiality of a left-linear rewrite system can be decided in
exponential time. Moreover, for left-linear rewrite systems satisfying
the additional syntactic condition that whenever two proper subterms of
left-hand sides are unifiable one of them matches the other, strong
sequentiality can be decided in polynomial time. The class of
forward-branching systems (Strandh~\cite{S89}), a proper subclass of
the class of orthogonal strongly sequential systems, coincides with the
class of transitive systems (Toyama~\textsl{et al.}~\cite{TSEP93}) and
can be decided in quadratic time (Durand~\cite{D94}). For classes
higher in the hierarchy no non-elementary upperbounds are known,
although Oyamaguchi~\cite{O98} believes that the time complexity of his
algorithm for deciding NV-sequentiality is at least double exponential.

In this paper we obtain a double exponential upperbound for the problem
of deciding whether a left-linear rewrite system belongs to $\CBNNFg$,
the largest class in the hierarchy of \cite{DM97}. This is better than
the complexity of the decision procedure in \cite{DM97} which, using
the analysis presented in this paper, is at least triple exponential in
the size of the rewrite system and much better than the non-elementary
upperbound that is obtained via the weak second-order monadic logic
connection.

The remainder of the paper is organized as follows. In the next section
we briefly recall the framework of our earlier paper \cite{DM97} for
analyzing call by need computations in term rewriting. In
Section~\ref{basic} we show that $(\to_\RR^*)[T]$ of ground terms that
rewrite to a term in $T$ is recognizable for every linear growing TRS
$\RR$ and recognizable $T$. This result, essentially originating from
\cite{C95} and \cite{J96}, forms the basis of the explicit construction
that we present in Section~\ref{CBNNF} of a tree automaton that decides
whether a left-linear TRS belongs to $\CBNNFg$. Section~\ref{CBNNF}
also contains an example illustrating the various constructions. The
complexity of the construction is analyzed in the next section. In
Section~\ref{CBNRS} we consider call by need computations to
root-stable form. As argued in Middeldorp~\cite{M97}, root-stable forms
and root-neededness are the proper generalizations of normal forms and
neededness when it comes to infinitary normalization. In this case we
again obtain a double exponential upperbound, which is a significant
improvement over the non-elementary upperbound of the complexity of the
decision procedure presented in \cite{DM97}.

\section{Preliminaries}

We assume the reader is familiar with the basics of term rewriting
(\cite{BN98,DJ90,K92}) and tree automata (\cite{TATA,GS84}). We recall
the following definitions from \cite{DM97}. We refer to the latter
paper for motivation and examples.

Let $\RR$ be a TRS over a signature $\FF$. The sets of ground redexes,
ground normal forms, and root-stable forms of $\RR$ are denoted by
$\REDEX_\RR$, $\NF_\RR$, and $\RS_\RR$. Let $\RR_\bullet$ be the TRS
$\RR \cup \{ \bullet \to \bullet \}$ over the extended signature
$\GG = \FF \cup \{ \bullet \}$. We say that redex $\Delta$ in
$C[\Delta] \in \TT(\FF)$ is $\RR$-needed if there is no term
$t \in \NF_{\RR_\bullet}$ such that $C[\bullet] \to_\RR^* t$. For
orthogonal TRSs $\RR$-neededness coincides with neededness.

Let $\RR$ and $\SS$ be TRSs over the same signature. We say that $\SS$
approximates $\RR$ if ${\to_\RR^*} \subseteq {\to_\SS^*}$ and
$\NF_\RR = \NF_\SS$. Next we define the approximations $\RRs$, $\RRnv$,
and $\RRg$ of a TRS $\RR$. The TRS $\RRs$ is obtained from $\RR$ by
replacing the right-hand side of every rewrite rule by a variable that
does not occur in the corresponding left-hand side. The TRS $\RRnv$ is
obtained from $\RR$ by replacing the variables in the right-hand sides
of the rewrite rules by pairwise distinct variables that do not occur
in the corresponding left-hand sides. A TRS $\RR$ is called growing if
for every rewrite rule $l \to r \in \RR$ the variables in
$\Var(l) \cap \Var(r)$ occur at depth $1$ in $l$. We define $\RRg$ as
any right-linear growing TRS that is obtained from $\RR$ by replacing
variables in the right-hand sides of the rewrite rules by variables
that do not occur in the corresponding left-hand sides.

An approximation mapping is a mapping $\alpha$ from TRSs to TRSs with
the property that $\alpha(\RR)$ approximates $\RR$, for every TRS
$\RR$. In the following we write $\RR_\alpha$ instead of $\alpha(\RR)$.
The class of TRSs $\RR$ such that every reducible term in $\TT(\FF)$
has an $\RR_\alpha$-needed redex is denoted by $\CBNNF_\alpha$. The
above definitions of $\RRs$, $\RRnv$, and $\RRg$ induce approximation
mappings $\mathrm{s}$, $\mathrm{nv}$, and $\mathrm{g}$. (We don't
consider shallow approximations in this paper since they are a
restricted case of growing approximations without resulting in any
obvious lower complexity.) It is known that $\CBNNFs$ coincides with
the class of strongly sequential TRSs but $\CBNNFnv$ and $\CBNNFg$
are much larger than the corresponding classes based on the
sequentiality concept, see \cite{DM97}.

Let $\RR$ be a TRS over a signature $\FF$. Let $\RR^\circ$ be the TRS
$\RR \cup \{ l^\circ \to r \mid l \to r \in \RR \}$ over the extended
signature $\GG = \FF \cup \{ f^\circ \mid f \in \FF \}$. Here $l^\circ$
denotes the term $f^\circ(\seq{l})$ for $l = f(\seq{l})$. For TRSs
$\RR$ and $\SS$ over the same signature $\FF$, we say that redex
$\Delta$ in $C[\Delta] \in \TT(\FF)$ is $(\RR,\SS)$-root-needed if
there is no term $t \in \RS_{\SS^\circ}$ such that
$C[\Delta^\circ] \to_\RR^* t$. Let $\alpha$ and $\beta$ be
approximation mappings. The class of TRSs $\RR$ such that every
non-$\RR_\beta$-root-stable term in $\TT(\FF)$ has an
$(\RR_\alpha,\RR_\beta)$-root-needed redex is denoted by
$\CBNRS_{\alpha,\beta}$.

\section{Basic Construction}
\label{basic}

We consider finite bottom-up tree automata without
$\epsilon$-transitions. Let $\RR$ be a linear growing TRS over a
signature $\FF$. Let $T \subseteq \TT(\GG)$ be a recognizable tree
language with $\FF \subseteq \GG$ and
$\AA_T = (\GG, Q_\AA, Q_f, \Gamma_\AA)$ a tree automaton that
recognizes $T$. We assume without loss of generality that all states of
$\AA_T$ are accessible.

The goal of this section is to construct a tree automaton that
recognizes the set $(\to_\RR^*)[T]$ of ground terms that rewrite to a
term in $T$. This result is not very new (Jacquemard~\cite{J96} gives
the construction for $T = \NF_\RR$ and Comon~\cite{C99}
for arbitrary $T$ and shallow $\RR$), but our presentation of the proof
is a bit crisper. More importantly, we need the details of the
construction for further analysis in subsequent sections.

\subsection{Step 1}
\label{step1}

Let $A_\RR$ be the set of arguments of the left-hand sides of $\RR$ and
let $S_\RR$ be the set of all subterms of terms in $A_\RR$. Construct
the tree automaton $\BB(\RR) = (\GG, Q_\BB, \varnothing, \Gamma_\BB)$
with $Q_\BB = \{ \state{t} \mid t \in S_\RR \} \cup \{ \state{x} \}$
and $\Gamma_\BB$ consisting of the \emph{matching} rules
$f(\state{t_1}, \dots, \state{t_n}) \to \state{t}$ for every term
$t = f(\seq{t})$ in $S_\RR$ and \emph{propagation} rules
$f(\state{x}, \dots, \state{x}) \to \state{x}$ for every $f \in \GG$.
Here $\state{t}$ denotes the equivalence class of the term $t$ with
respect to literal similarity. (So we identify $\state{s}$ and
$\state{t}$ whenever $s$ and $t$ differ a variable renaming.) Note that
all states of $\BB(\RR)$ are accessible. From now on we write $\BB$ for
$\BB(\RR)$ when the TRS $\RR$ can be inferred from the context. The set
of ground instances of a term $t$ is denoted by $\Sigma(t)$.

\begin{lemma}
\label{basic_B}
Let $t \in S_\RR \cup \{ x \}$. We have $s \in \Sigma(t)$ if and only
if $s \to_\BB^* \state{t}$.
\qed
\end{lemma}

\subsection{Step 2}
\label{step2}

We assume that 
$\{ t \mid t \to_\AA^+ q \} = \{ t \mid t \to_\BB^+ q \}$ for every
state $q \in Q_\AA \cap Q_\BB$. This can always be achieved by a
renaming of states. Let $\CC_T(\RR) = (\GG, Q, Q_f, \Gamma)$ be the
union of $\AA_T$ and $\BB(\RR)$, so $Q = Q_\AA \cup Q_\BB$ and
$\Gamma = \Gamma_\AA \cup \Gamma_\BB$.

\begin{lemma}
\label{basic_A+B}
\mbox{}
\begin{enumerate}
\item
Let $t \in S_\RR \cup \{ x \}$. We have $s \in \Sigma(t)$ if and only
if $s \to_{\CC_T(\RR)}^* \state{t}$.
\item
$L(\CC_T(\RR)) = T$.
\end{enumerate}
\qed
\end{lemma}

\subsection{Step 3}
\label{step3}

We saturate the transition rules $\Gamma$ of $\CC_T(\RR)$ under the
following inference rule:
\begin{equation}
\frac
{f(\seq{l}) \to r \in \RR \quad r\theta \to_\Gamma^* q}
{\Gamma = \Gamma \cup \{f(\seq{q}) \to q\}}
\tag{$\ast$}
\end{equation}
with $\theta$ mapping the variables in $r$ to states in $Q$ and
\[
q_i = \begin{cases}
l_i\theta & \text{if $l_i \in \Var(r)$,} \\
\state{l_i} & \text{otherwise.}
\end{cases}
\]
Because $Q$ is finite and no new state is added by ($\ast$), the
saturation process terminates. We claim that
$L(\CC_T(\RR)) = (\to_\RR^*)[T]$ upon termination.

\begin{lemma}
\label{property1}
$(\to_\RR^*)[T] \subseteq L(\CC_T(\RR))$.
\end{lemma}
\begin{proof}
Let $s \in (\to_\RR^*)[T]$. So there exists a term $t \in T$ such that
$s \to_\RR^* t$. We show that $s \in L(\CC_T(\RR))$ by induction on the
length of $s \to_\RR^* t$. If $s = t$ then
$s \in T \subseteq L(\CC_T(\RR))$ according to
Lemma~\ref{basic_A+B}(2). Let
$s = C[l\sigma] \to_\RR C[r\sigma] \to_\RR^* t$ with $l = f(\seq{l})$.
The induction hypothesis yields $C[r\sigma] \in L(\CC_T(\RR))$. Hence
there exists a final state $q_f$, a mapping $\theta$ from $\Var(r)$ to
$Q$, and a state $q$ such that
$C[r\sigma] \to_\Gamma^* C[r\theta] \to_\Gamma^* C[q] \to_\Gamma^* q_f$.
By construction there exists a transition rule
$f(\seq{q}) \to q \in \Gamma$ such that $q_i = l_i\theta$ if
$l_i \in \Var(r)$ and $q_i = \state{l_i}$ otherwise. We claim that
$l\sigma \to_\Gamma^* f(\seq{q})$. Let $i \in \{ 1, \ldots, n \}$. If
$l_i \in \Var(r)$ then $l_i\sigma \to_\Gamma^* l_i\theta = q_i$,
otherwise $l_i\sigma \to_\Gamma^* \state{l_i} = q_i$ by
Lemma~\ref{basic_A+B}(1). Consequently
$s \to_\Gamma^* C[f(\seq{q})] \to_\Gamma C[q] \to_\Gamma^* q_f$
and hence $s \in L(\CC_T(\RR))$.
\qed
\end{proof}

Note that we don't use the growing assumption in the above
proof; right-linearity of $\RR$ is sufficient.

\begin{lemma}
\label{property2}
Let $s \in \TT(\GG)$ with $s \to_{\CC_T(\RR)}^* q$.
\begin{enumerate}
\item
If $q = \state{t}$ with $t \in S_\RR \cup \{ x \}$ then
$s \in (\to_\RR^*)[\Sigma(t)]$.
\item
If $q \in Q_f$ then $s \in (\to_\RR^*)[T]$.
\end{enumerate}
\end{lemma}
\begin{proof}
Let $\Gamma_k$ denote the value of $\Gamma$ after the $k$-th transition
rule has been added by ($\ast$). We have $s \to_{\Gamma_k}^* q$ for
some $k \geqslant 0$. We prove statements (1) and (2) by induction on
$k$. If $k = 0$ then the result follows from Lemma~\ref{basic_A+B}. Let
$s \to_{\Gamma_{k+1}}^* q$. We use a second induction on the number of
steps that use the (unique) transition rule
$f(\seq{q}) \to q' \in \Gamma_{k+1} \setminus \Gamma_k$. Suppose this
rule is created from $l = f(\seq{l}) \to r \in \RR$ and
$r\theta \to_{\Gamma_k}^* q'$. If this number is zero then the result
follows from the first induction hypothesis. Otherwise we may write
$s = C[f(\seq{s})] \to_{\Gamma_k}^* C[f(\seq{q})] \to C[q']
\to_{\Gamma_{k+1}}^* q$.
We will define a substitution $\tau$ such that
$s \to_\RR^* C[l\tau] \to_\RR C[r\tau] \to_{\Gamma_k}^+ C[q']$.
The second induction hypothesis applied to
$C[r\tau] \to_{\Gamma_{k+1}}^+ q$ then yields the desired result. We
define $\tau$ as the (disjoint) union of $\tau_1, \dots, \tau_n, \tau'$
such that $\Dom(\tau_i) = \Var(l_i)$ for $i = 1, \dots, n$ and
$\Dom(\tau') = \Var(r) \setminus \Var(l)$. Note that since $l$ is a
linear term, the union of $\tau_1, \dots, \tau_n$ is well-defined. Fix
$i \in \{ 1, \dots, n \}$. If $l_i \in \Var(r)$ then we let
$\tau_i = \{ l_i \mapsto s_i \}$. Otherwise $q_i = \state{l_i}$ and
thus $s_i \to_{\Gamma_k}^+ \state{l_i}$. Part (1) of the first
induction hypothesis yields $s_i \in (\to_\RR^*)[\Sigma(l_i)]$. Hence
there exists a substitution $\tau_i$ such that
$s_i \to_\RR^* l_i\tau_i$. We assume without loss of generality that
$\Dom(\tau_i) = \Var(l_i)$. The substitution $\tau'$ is defined as
$\{ x \mapsto u_x \mid x \in \Var(r) \setminus \Var(l) \}$ where $u_x$
is an arbitrary but fixed ground term such that
$u_x \to_{\Gamma_0}^* x\theta$. (This is possible because all states of
$Q$ are accessible.) It remains to show that $s \to_\RR^* C[l\tau]$ and
$C[r\tau] \to_{\Gamma_k}^+ C[q']$. The former is an immediate
consequence of the definitions of $\seq{\tau}$. For the latter it is
sufficient to show that $C[r\tau] \to_{\Gamma_k}^* C[r\theta]$. Let
$x \in \Var(r)$. If $x \in \Var(l)$ then, because $\RR$ is growing and
left-linear, there is a unique $i \in \{ 1, \dots, n \}$ such that
$x = l_i$. We have $x\tau = l_i\tau_i = s_i$ by construction of
$\tau_i$ and $q_i = l_i\theta = x\theta$ by definition. Hence
$x\tau = s_i \to_{\Gamma_k}^+ q_i = x\theta$ by assumption. If
$x \notin \Var(l)$ then $x\tau = x\tau' \to_{\Gamma_0}^* x\theta$ by
construction of $\tau'$. This completes the proof. The induction step
is summarized in the following diagram:
\[
\xymatrix{
s \ar[r] _-{\Gamma_k} ^-{*} \ar[d] _-{\RR} ^-{*} &
C[f(\seq{q})] \ar[r] _-{\Gamma_{k+1}} &
C[q'] \ar[r] _-{\Gamma_{k+1}} ^-{*} &
q \\
C[l\tau] \ar[r] _-{\RR} &
C[r\tau] \ar[r] _-{\Gamma_k} ^-{*} &
C[r\theta] \ar[u] _-{\Gamma_k} ^-{*}
}
\]
\qed
\end{proof}

\begin{corollary}
$L(\CC_T(\RR)) = (\to_\RR^*)[T]$.
\qed
\end{corollary}

As a side remark we mention that the result described above remains
true if we drop the restriction that the left-hand side of a rewrite
rule is a non-variable term; just add the following saturation rule:
\begin{equation}
\frac
{x \to r \in \RR \quad r\theta \to_\Gamma^* q}
{\Gamma = \Gamma \cup \{q' \to q\}}
\tag{$\ast\ast$}
\end{equation}
with $\theta$ mapping the variables in $r$ to states in $Q$ and
\[
q' = \begin{cases}
x\theta & \text{if $x \in \Var(r)$,} \\
\state{x} & \text{otherwise.}
\end{cases}
\]
Although this extension is useless when it comes to call-by-need
(because no TRS that has a rewrite rule whose left-hand side is a
single variable has normal forms), it is interesting to note that it
generalizes Theorem~5.1 of Coquid\'e \textsl{et al.}
\cite{CDGV94}---the preservation of recognizability for linear
semi-monadic rewrite systems.

For an example of the above constructions we refer to the end of the
next section.

\section{Call by Need Computations to Normal Form}
\label{CBNNF}

In this section we assume that $\GG = \FF \cup \{ \bullet \}$ and
$T = \NF_{\RR_\bullet} \subseteq \TT(\FF)$, the set of ground normal
forms of the linear growing TRS $\RR$. Based on the automaton
$\CC_T(\RR)$ of the previous section we construct an automaton
$\DD(\RR)$ (or simply $\DD$) that accepts all reducible terms in
$\TT(\FF)$ that do not have an $\RR$-needed redex.

First we extend $\CC_T(\RR)$ in such a way that it can be used to
identify redexes and reducible terms with respect to $\RR$. This is
essentially achieved by adding a fresh copy of the automaton $\BB(\RR)$
of Section~\ref{step1}. More precisely, we add for every term
$t \in S_\RR$ a new state $\new{t}$ to $Q$ as well as a distinguished
state $q_r$ together with the transition rules
$f(\new{t_1}, \dots, \new{t_n}) \to \new{t}$ for every term
$t = f(\seq{t}) \in S_\RR$ and $f(\new{l_1}, \dots, \new{l_n}) \to q_r$
for every left-hand side $f(\seq{l})$ of a rewrite rule in $\RR$, where
we identify $\new{x}$ with $\state{x}$, and all rules of the form
$f(\state{x}, \dots, q_r, \dots, \state{x}) \to q_r$. In state $q_r$
all reducible terms in $\TT(\GG)$ are accepted. Let us denote the new
automaton by $\CC'_T(\RR) = (\GG, Q', Q_f, \Gamma')$. Note that
$\CC_T(\RR)$ and $\CC'_T(\RR)$ accept the same language
$(\to_\RR^*)[T]$.

The construction of $\DD$ is related to the construction in
Comon~\cite[Lemma~31]{C99}. The signature of $\DD$ is
$\FF$. The set $Q_\DD$ of states of $\DD$ consists of all pairs $[S,P]$
of subsets $S$, $P$ of states in $Q'$. The final states all are pairs
$[S,P]$ such that $q_r \in S$ and $P \subseteq Q_f$. Intuitively, the
first component $S$ of a state $[S,P]$ records all reachable states
with respect to the automaton $\CC'_T(\RR)$ and the second component
$P$ will be a subset of $S$ that can be divided in non-empty sets $P_p$
for every redex position $p$ such that all states in $P_p$ are
reachable if the redex at position $p$ is not contracted. This will be
made precise in Lemmata~\ref{final-P1} and~\ref{final-P2} below.

Before defining the transition rules of $\DD$ we introduce the
following abbreviations. Given a term $t \in \TT(\GG \cup Q_\DD)$, let
$Q(t)$ denote the set of all terms in $\TT(\GG \cup Q')$ that are
obtained from $t$ by replacing every state in $Q_\DD$ by one of its
elements. We denote the set
$\{ q \in Q' \mid \text{$s \to_{\Gamma'}^* q$ and $s \in Q(t)$} \}$ by
$t{\downarrow}$. The set $\Gamma_\DD$ consists of all transition rules
of the form
\[
f([S_1,P_1], \dots, [S_n,P_n]) \to [S,P]
\]
where $f$ is an $n$-ary function symbol in $\FF$,
$S = f(\seq{S}){\downarrow}$,
and $P$ is the union of $P^1$ and $P^2$ where $P^1$ is a subset of
\[
\bigcup_{i=1}^n f(S_1, \dots, P_i, \dots, S_n){\downarrow}
\]
with the property that for all $i \in \{ 1, \dots, n \}$ and
$q_i \in P_i$
\[
P^1 \cap f(S_1, \dots, \{ q_i \}, \dots, S_n){\downarrow} \neq
\varnothing
\]
and
$P^2 = \{ \state{x} \}$
if there exists a rewrite rule in $\RR$ with left-hand side
$f(\seq{l})$ such that $\new{l_i} \in S_i$ for all
$i \in \{ 1, \dots, n \}$ and
$P^2 = \varnothing$
otherwise. Note that the resulting automaton is non-deterministic due
to the freedom of choosing $P^1$. This is essential for the correctness
of the construction.

\begin{lemma}
\label{final-S}
Let $s \in \TT(\FF)$. If $s \to_{\Gamma_\DD}^* [S,P]$ then
$S = s{\downarrow}$.
\end{lemma}
\begin{proof}
Straightforward.
\qed
\end{proof}

\begin{lemma}
\label{final-P1}
Let $s \in \TT(\FF)$. If $s \to_{\Gamma_\DD}^* [S,P]$ then for every
redex position $p$ in $s$ there exists a state $q \in P$ such that
$q \in s[\bullet]_p{\downarrow}$.
\end{lemma}
\begin{proof}
Let $s_{|p} = l\sigma$ for some left-hand side $l = f(\seq{l})$ of a
rewrite rule in $\RR$. We have $l_i\sigma \to_{\Gamma'}^* \new{l_i}$
for all $i \in \{ 1, \dots, n \}$. We show the statement by induction
on the depth of the position $p$. If $p = \varepsilon$ then
$s = l\sigma$. We may write
\[
s = f(l_1\sigma, \dots, l_n\sigma) \to_{\Gamma_\DD}^*
f([S_1,P_1], \ldots, [S_n,P_n]) \to_{\Gamma_\DD} [S,P].
\]
According to Lemma~\ref{final-S} $\new{l_i} \in S_i$ for all
$i \in \{ 1, \dots, n \}$. By construction $P = P^1 \cup P^2$ for some
$P^1$ and $P^2 = \{ \state{x} \}$. It is sufficient to show that
$s[\bullet]_p = \bullet \to_{\Gamma'}^* \state{x}$, which is obvious
since by construction every ground term in $\TT(\GG)$ can be rewritten
to $\state{x}$. For the induction step we suppose that
$p = i {\cdot} p'$ for some $i \in \{ 1, \dots, n \}$. Write
$s = f(s_1, \dots, s_n) \to_{\Gamma_\DD}^*
f([S_1,P_1], \ldots, [S_n,P_n]) \to_{\Gamma_\DD} [S,P]$.
The induction hypothesis yields $q' \in s_i[\bullet]_{p'}{\downarrow}$
for some $q' \in P_i$. By construction $P = P^1 \cup P^2$ where
$P^1$ has a non-empty intersection with
$f(S_1, \dots, \{ q' \}, \dots, S_n){\downarrow}$.
Let $q$ be a state in this intersection. By definition there exist
states $q_j$ for $j \neq i$ such that
$f(q_1, \dots, q', \dots, q_n) \to_{\Gamma'} q$. Lemma~\ref{final-S}
yields $s_j \to_{\Gamma'}^* q_j$ for all $j \neq i$. Hence
$s[\bullet]_p = f(s_1, \dots, s_i[\bullet]_{p'}, \dots, s_n)
\to_{\Gamma'}^* f(q_1, \dots, q', \dots, q_n)
\to_{\Gamma'} q$
and thus $q \in s[\bullet]_p{\downarrow}$.
\qed
\end{proof}

We denote the set of redex positions in a term $s$ by $\RR(s)$.

\begin{lemma}
\label{final-P2}
Let $s \in \TT(\FF)$. If
\[
P \subseteq \bigcup_{p \in \RR(s)} s[\bullet]_p{\downarrow}
\]
such that $P \cap s[\bullet]_p{\downarrow} \neq \varnothing$ for all
$p \in \RR(s)$ then $s \to_{\Gamma_\DD}^* [S,P]$ where
$S = s{\downarrow}$.
\end{lemma}
\begin{proof}
Induction on the structure of $s$. Suppose $s = f(\seq{s})$ for some
$n \geqslant 0$. Define
\[
P' = P \cap
\bigcup_{\makebox[1cm]{$\scriptstyle \varepsilon \neq p \in \RR(s)$}}
s[\bullet]_p{\downarrow}.
\]
Fix $i \in \{ 1, \dots, n \}$ and define
\[
P_i = \bigcup_{p' \in \RR(s_i)} 
\{ \text{$q' \in s_i[\bullet]_{p'}{\downarrow}$ such that
$f(S_1, \dots, \{ q' \}, \dots, S_n){\downarrow} \cap P' \neq
\varnothing$}
\}.
\]
Clearly
\[
P_i \subseteq \bigcup_{p' \in \RR(s_i)} s_i[\bullet]_{p'}{\downarrow}.
\]
We claim that $P_i \cap s_i[\bullet]_{p'}{\downarrow} \neq \varnothing$
for all $p' \in \RR(s_i)$. Let $p = i {\cdot} p'$. By assumption
$P \cap s[\bullet]_p{\downarrow} \neq \varnothing$. We have
$s[\bullet]_p{\downarrow} =
f(S_1, \dots, s_i[\bullet]_{p'}{\downarrow}, \dots, S_n){\downarrow}$,
so there exists a state $q' \in s_i[\bullet]_{p'}{\downarrow}$ such
that $f(S_1, \dots, \{ q' \}, \dots, S_n){\downarrow} \cap P \neq
\varnothing$. This implies that
$f(S_1, \dots, \{ q' \}, \dots, S_n){\downarrow} \cap P' \neq
\varnothing$. Hence $q' \in P_i$ by definition. The induction
hypothesis yields $s_i \to_{\Gamma_\DD}^* [S_i,P_i]$ with
$S_i = s_i{\downarrow}$. Since this holds for every
$i \in \{ 1, \dots, n \}$ we obtain
\[
s \to_{\Gamma_\DD}^* f([S_1,P_1], \dots, [S_n,P_n]).
\]
It suffices to show that
$f([S_1,P_1], \dots, [S_n,P_n]) \to [S,P]$ is a transition rule of
$\DD$, which amounts to showing that $S = f(\seq{S}){\downarrow}$ and
$P = P^1 \cup P^2$ with $P^1$ a subset of
\[
\bigcup_{i=1}^n f(S_1, \dots, P_i, \dots, S_n){\downarrow}
\]
with the property that for all $i \in \{ 1, \dots, n \}$ and
$q_i \in P_i$
\[
P^1 \cap f(S_1, \dots, \{ q_i \}, \dots, S_n){\downarrow} \neq
\varnothing
\]
and $P^2 = \{ \state{x} \}$ if there exists a rewrite rule in $\RR$
with left-hand side $f(\seq{l})$ such that
$\new{l_1} \in S_1, \dots, \new{l_n} \in S_n$ and $P^2 = \varnothing$
otherwise. Clearly
\[
f(\seq{S}){\downarrow} =
f(s_1{\downarrow}, \dots, s_n{\downarrow}){\downarrow} =
s{\downarrow} = S.
\]
For $P^1$ we take $P'$. By the definition of $\seq{P}$, for all
$i \in \{ 1, \dots, n \}$ and $q_i \in P_i$
\[
P' \cap f(S_1, \dots, \{ q_i \}, \dots, S_n){\downarrow} \neq
\varnothing,
\]
so $P'$ satisfies the conditions for $P^1$. If $s$ is a redex then
there exist a left-hand side $f(\seq{l})$ of a rewrite rule in $\RR$
and a substitution $\sigma$ such that
$s = f(l_1\sigma, \dots, l_n\sigma)$. We have
$P = P' \cup \bullet{\downarrow} = P' \cup \{ \state{x} \}$ and
$P^2 = \{ \state{x} \}$ as desired since according to
Lemma~\ref{final-S} $\new{l_i} \in S_i$ for all
$i \in \{ 1, \dots, n \}$. If $s$ is not a redex then $P' = P$ and
indeed $P^2 = \varnothing$ in this case since there does not exist a
left-hand side $f(\seq{l})$ such that $\new{l_i} \in S_i$ for all
$i \in \{ 1, \dots, n \}$.
\qed
\end{proof}

\begin{lemma}
\label{final-P3}
Let $s \in \TT(\FF)$. If $s \to_{\Gamma_\DD}^* [S,P]$ and
$P \subseteq Q_f$ then, for every redex position $p$ in $s$,
$s[\bullet]_p \in (\to_\RR^+)[\NF_{\RR_\bullet}]$.
\end{lemma}
\begin{proof}
Let $p$ be a redex position in $s$. By Lemma~\ref{final-P1} there
exists a state $q \in Q_f$ such that $s[\bullet]_p \to_{\Gamma'}^* q$.
Hence $s[\bullet]_p \in L(\CC'_{\NF_{\RR_\bullet}}) =
(\to_\RR^*)[\NF_{\RR_\bullet}]$.
\qed.
\end{proof}

\begin{lemma}
\label{final-P4}
Let $s \in \TT(\FF)$. If
$s[\bullet]_p \in (\to_\RR^*)[\NF_{\RR_\bullet}]$ for every redex
position $p$ in $s$ then there exists a set $P \subseteq Q_f$ such
that $s \to_{\Gamma_\DD}^* [S,P]$ with $S = s{\downarrow}$.
\end{lemma}
\begin{proof}
Define
\[
P = Q_f \cap \bigcup_{p \in \RR(s)} s[\bullet]_p{\downarrow}.
\]
Since $(\to_\RR^*)[\NF_{\RR_\bullet}] = L(\CC'_{\NF_{\RR_\bullet}})$,
for every redex position $p$ in $s$ there exists a state $q \in Q_f$
such that $s[\bullet]_p \to_{\Gamma'}^* q$ and thus $q \in P$ by
definition. Hence we can apply Lemma~\ref{final-P2} which yields the
desired result.
\qed
\end{proof}

\begin{corollary}
Let $s \in \TT(\FF)$. We have $s \in L(\DD(\RR))$ if and only if $s$
is reducible and $s[\bullet]_p \in (\to_\RR^*)[\NF_{\RR_\bullet}]$ for
every redex position $p$ in $s$.
\end{corollary}
\begin{proof}
By Lemmata~\ref{final-S}, \ref{final-P3}, \ref{final-P4}, and the
observation that $s$ is reducible if and only if
$q_r \in s{\downarrow}$.
\qed
\end{proof}

\begin{theorem}
Let $\RR$ be a left-linear TRS. We have $\RR \in \CBNNFg$ if and only
if $L(\DD(\RRg)) = \varnothing$.
\end{theorem}
\begin{proof}
Note that $\RR_g$ is a linear growing TRS. By definition of $\CBNNFg$,
$\RR \notin \CBNNFg$ if and only if there exists a reducible term $s$
in $\TT(\FF)$ without $\RRg$-needed redexes. The latter is equivalent
to $s[\bullet]_p \in (\to_{\RRg}^*)[\NF_{\RR_\bullet}]$ for every redex
position $p$ in $s$. (Note that $\NF_{\RR_\bullet}$ and
$\NF_{{\RRg}_\bullet}$ coincide.) According to the preceding corollary
this is equivalent to $s \in L(\DD(\RRg))$. Hence $\RR \in \CBNNFg$ if
and only if $L(\DD(\RRg)) = \varnothing$.
\qed
\end{proof}

Let us illustrate the construction on a small example. Consider
the orthogonal growing TRS $\RR$ consisting of the rewrite rules
\[
\begin{array}{rcl@{\qquad}rcl}
f(a,g(x,a)) & \to & b &
f(x,a)      & \to & x \\
f(b,g(a,x)) & \to & b &
g(b,b)      & \to & a
\end{array}
\]
The automaton $\BB(\RR)$ has states $Q_\BB =
\{ \state{x}, \state{a}, \state{b}, \state{g(x,a)}, \state{g(a,x)} \}$
and transition rules $\Gamma_\BB$:
\[
\begin{array}{rcl@{\qquad}rcl@{\qquad}rcl@{\qquad}rcl}
a & \to & \state{x} &
\bullet & \to & \state{x} &
a & \to & \state{a} &
g(\state{x},\state{a}) & \to & \state{g(x,a)} \\
b & \to & \state{x} &
f(\state{x},\state{x}) & \to & \state{x} &
b & \to & \state{b} &
g(\state{a},\state{x}) & \to & \state{g(a,x)}
\end{array}
\]
The set $\NF_{\RR_\bullet}$ is for instance accepted by the automaton
$\AA_{\NF_{\RR_\bullet}}$ with states
$Q_\AA = Q_f = \{ \state{X}, \state{a}, \state{b}, \state{G(X,a)},
\state{G(a,X)}, \state{G(a,a)} \}$ and transition rules $\Gamma_\AA$
consisting of $a \to \state{a}$, $b \to \state{b}$,
\[
\begin{array}{rcl}
g(q,q') & \to &
\begin{cases}
\state{G(a,a)} & \mbox{if $q = q' = \state{a}$} \\
\state{G(a,X)} & \mbox{if $q = \state{a}$ and
$q' \in Q_\AA \setminus \{ \state{a} \}$} \\
\state{G(X,a)} & \mbox{if $q \in Q_\AA \setminus \{ \state{a} \}$ and
$q' = \state{a}$} \\
\state{X} & \mbox{for all other cases except $q = q' = \state{b}$}
\end{cases}
\end{array}
\]
and $f(q,q') \to \state{X}$ for all pairs $(q,q')$ except those in
$Q_\AA \times \{ \state{a} \}$,
$\{ \state{a} \} \times \{ \state{G(X,a)}, \state{G(a,a)} \}$, and
$\{ \state{b} \} \times \{ \state{G(a,X)}, \state{G(a,a)} \}$.
Note that $\AA_{\NF_{\RR_\bullet}}$ and $\BB(\RR)$ share states
$\state{a}$ and $\state{b}$, which is allowed since both automata
accept the same set of terms in those states ($\{ a \}$ and $\{ b \}$
respectively). Let $Q = Q_\AA \cup Q_\BB$ and
$\Gamma = \Gamma_\AA \cup \Gamma_\BB$. Let us compute the saturation
rules for the various approximations of $\RR$. For $\RRs$ we get
\[
\begin{array}{rcl@{\qquad}rcl}
f(\state{a},\state{g(x,a)}) & \to & q &
f(\state{x},\state{a}) & \to & q \\
f(\state{b},\state{g(a,x)}) & \to & q &
g(\state{b},\state{b}) & \to & q
\end{array}
\]
for all states $q \in Q$. For $\RRnv$ we obtain
\[
\begin{array}{rcl@{\qquad}rcl}
f(\state{a},\state{g(x,a)}) & \to & q &
f(\state{x},\state{a}) & \to & q_1 \\
f(\state{b},\state{g(a,x)}) & \to & q &
g(\state{b},\state{b}) & \to & q_2
\end{array}
\]
for all $q \in \{ \state{x}, \state{b} \}$, $q_1 \in Q$, and
$q_2 \in \{ \state{x}, \state{a} \}$. Finally, $\RRg$ gives rise to
\[
\begin{array}{rcl@{\qquad}rcl}
f(\state{a},\state{g(x,a)}) & \to & q &
f(q_1,\state{a}) & \to & q_1 \\
f(\state{b},\state{g(a,x)}) & \to & q &
g(\state{b},\state{b}) & \to & q_2
\end{array}
\]
for all $q \in \{ \state{x}, \state{b} \}$, $q_1 \in Q$, and
$q_2 \in \{ \state{x}, \state{a} \}$. Let us denote the resulting
set of transition rules of the automaton
$\CC_{\NF_{\RR_\bullet}}(\RR_\alpha)$ for
$\alpha \in \{ \mathrm{s}, \mathrm{nv}, \mathrm{g} \}$ by
$\Gamma_\alpha$. Note the difference between $\Gamma_\mathrm{nv}$ and
$\Gamma_\mathrm{g}$.
We obtain $\CC'_{\NF_{\RR_\bullet}}(\RR_\alpha)$ from
$\CC_{\NF_{\RR_\bullet}}(\RR_\alpha)$ by adding states
$\{ \state{a}', \state{b}', \state{g(x,a)}', \state{g(a,x)}', q_r \}$
and transition rules
\[
\begin{array}{rcl@{\qquad}rcl@{\qquad}rcl}
a & \to & \state{a}' &
f(\state{a}',\state{g(x,a)}') & \to & q_r &
f(\state{x},q_r) & \to & q_r \\
b & \to & \state{b}' &
f(\state{b}',\state{g(a,x)}') & \to & q_r &
f(q_r,\state{x}) & \to & q_r \\
g(\state{x},\state{a}') & \to & \state{g(x,a)}' &
f(\state{x},\state{a}') & \to & q_r &
g(\state{x},q_r) & \to & q_r \\
g(\state{a}',\state{x}) & \to & \state{g(a,x)}' &
g(\state{b}',\state{b}') & \to & q_r &
g(q_r,\state{x}) & \to & q_r
\end{array}
\]
We will not attempt to present the automata $\DD(\RR_\alpha)$ in
detail. Rather, we show a possible computation of $\DD(\RRnv)$ for the
term $t = f(\Delta,g(\Delta,\Delta))$ with $\Delta = g(a,a)$. We have
$a \to [\{ \state{x}, \state{a}, \state{a}' \}, \varnothing]$ and thus
\[
\Delta \to^* f([\{ \state{x}, \state{a}, \state{a}' \},\varnothing],
[\{ \state{x}, \state{a}, \state{a}' \}, \varnothing]) \to
[S_1,P_1]
\]
with
$S_1 = Q' \setminus \{ \state{b}', \state{g(x,a)}', \state{g(a,x)}' \}$
and $P_1 = \{ \state{x} \}$. Consequently,
\[
g(\Delta,\Delta) \to^* g([S_1,P_1],[S_1,P_1]) \to [S_2,P_2]
\]
with $S_2 = \{ \state{x}, \state{a}, \state{g(x,a)}, \state{g(a,x)},
\state{X}, \state{G(X,a)}, \state{G(a,X)}, \state{G(a,a)}, q_r \}$ and
$P_2 = \{ \state{g(x,a)}, \state{g(a,x)} \}$. Note that there are other
possibilities for $P_2$. Finally
\[
t \to^* f([S_1,P_1],[S_2,P_2]) \to [S,P]
\]
with $S = t{\downarrow_{\Gamma_\mathrm{nv}'}}$ and
$P = \{ \state{X} \}$. Since $q_r \in S$ and $\state{X} \in Q_f$,
$[S,P]$ is a final state of $\DD(\RRnv)$ and hence $t$ does not have an
$\RRnv$-needed redex and thus $\RR \notin \CBNNFnv$. The reader is
invited to verify that for every computation $t \to^* [S,P]$ in
$\DD(\RRg)$ we have $\state{x} \in P$ and thus $t$ has an $\RRg$-needed
redex (the occurrence of $\Delta$ at position $1$). Actually, it turns
out that $L(\DD(\RRg)) = \varnothing$, so $\RR \in \CBNNFg$.

\section{Complexity Analysis}
\label{complexity}

In this section we analyze the complexity of the decision procedure of
the previous section. Given a term $t$, we denote its size (i.e., its
total number of symbols) by $|t|$. Given a TRS $\RR$, we denote the
number of rewrite rules it contains by $\sharp\RR$ and its size (the
sum of the sizes of the left and right-hand sides) by $|\RR|$. We
assume that the signature $\FF$ of a $\RR$ does not contain function
symbols that do not appear in $\RR$, except possibly for a constant
(to make the set $\TT(\FF)$ of ground terms non-empty). This entails
no loss of generality. Let $m$ be the maximum arity of function
symbols in $\FF$.

Given an automaton $\AA$, we denote the number of transitions rules of
$\AA$ by $|\AA|$. It is well-known that the number of states of an
automaton $\AA_{\NF_{\RR_\bullet}}$ that accepts the ground normal
forms in $\TT(\FF)$ with respect to the TRS $\RR$ is in
$\OO(2^{|\RR|})$. Hence $|\AA|$ (from now on we drop the subscript
$\NF_{\RR_\bullet}$) is in $\OO(2^{\OO(|\RR|^2)})$: For every function
symbol in $\FF$ there can be at most $|Q_\AA|^{m+1}$ transition rules
hence $|\AA|$ is in $\OO(|\FF| \cdot 2^{|\RR|(m+1)})$ and thus in
$\OO(2^{\OO(|\RR|^2)})$ since we can estimate $|\FF|$ and $m$ by
$|\RR|$. Compared to $\AA$ the size of the automaton $\BB(\RR)$ is
neglectable. Next we analyze the complexity of the saturation process
of Sect.~\ref{step3}. (A similar analysis is reported in
\cite[chapitre IV]{J96PhD}.)

\begin{lemma}
\label{complexity-saturation}
Let $\RR$ be a growing TRS. The saturation rules of $\CC_T(\RR)$ can be
computed in $\OO(|\RR|^5 \cdot |Q|^{\OO(|\RR|)})$ time.
\end{lemma}
\begin{proof}
Let $\Xi$ be the set of transition rules that may potentially appear in
in the automaton $\CC_T(\RR)$: $\Xi = \{ f(\seq{q}) \to q \mid
\mbox{$f \in \FF$ and $\seq{q}, q \in Q$} \}$. There are at most
$K = \sharp\RR \cdot |Q|^{m+1}$ rules in $\Xi$.
The saturation process may be described by the following algorithm:
\begin{quote}
\small
\begin{tabbing}
$\Xi_0 := \Xi \setminus \Gamma_0$; \\
$k := 1$; \\
\texttt{while} \\
\mbox{}\qquad \= $\begin{array}{ll}
\exists\,f(\seq{q}) \to q \in \Xi_{k-1} \\
\exists\,f(\seq{l}) \to r \in \RR & \mbox{~such that~} \\ 
\exists\,\theta\colon \Var(r) \to Q
\end{array}$
\parbox{5cm}{
$r\theta \to_{\Gamma_{k-1}}^* q$ and,
for all $1 \leqslant i \leqslant n$,
$q_i = l_i\theta$ if $l_i \in \Var(r)$ and
$q_i = \state{l_i}$ otherwise
} \\
\texttt{do} \\
\> $\Xi_k := \Xi_{k-1} \setminus \{ f(\seq{q}) \to q\}$; \\
\> $\Gamma_k := \Gamma_{k-1} \cup \{ f(\seq{q}) \to q \}$; \\
\> $k := k + 1$
\end{tabbing}
\end{quote}
Let us estimate the time to evaluate the condition of the while-loop.
There are $K - |\Gamma_{k-1}|$ choices for $f(\seq{q}) \to q$,
$\sharp\RR$ choices for $f(\seq{l}) \to r$, and $|Q|^{\Var(r)}$ choices
for $\theta$. For every choice we have to test whether
$r\theta \to_{\Gamma_{k-1}}^* q$ is true. (The other requirements are
neglectable.) This can be done in $\OO(|r| \cdot |\Gamma_{k-1}|)$ time.
So one iteration of the while-loop takes
\[
\OO((K - |\Gamma_{k-1}|) \cdot \sharp\RR \cdot |Q|^{\Var(r)} \cdot
|r| \cdot |\Gamma_{k-1}|)
\]
time. To obtain the time complexity of the algorithm we have to
multiply this by the maximum number of iterations, which is
$K - |\Gamma_0|$. Removing the negative terms and estimating
$|\Gamma_{k-1}|$ by $K$ and $\Var(r)$ by $|r|$ yields
\[
\OO(K^3 \cdot \sharp\RR \cdot |Q|^{|r|} \cdot |r|)
=
\OO(\sharp\RR^4 \cdot |Q|^{3m+3+|r|} \cdot |r|)
\]
Estimating $\sharp\RR$, $m$, and $|r|$ by $|\RR|$ yields the complexity
class $\OO(|\RR|^5 \cdot |Q|^{\OO(|\RR|)})$ in the statement of the
lemma. 
\qed
\end{proof}

So for growing TRSs the time complexity of the saturation process is
exponential in the size of the TRS. For $\RRs$ and $\RRnv$ we get a
polynomial time complexity, but the space and time complexity of the
automaton $\CC$ is still exponential in $|\RR|$ due to the normal form
automaton.

The number $|Q|$ of states of the automaton $\CC(\RRg)$ is of the same
order as the number of states of $\AA$. The time to compute $\CC(\RRg)$
is the sum of the times to compute the automaton $\AA$
($\OO(2^{\OO(|\RR|^2)})$), the automaton $\BB(\RR)$ (neglectable), and
the saturation rules ($\OO(2^{\OO(|\RR|^2)})$ from
Lemma~\ref{complexity-saturation} with $|Q| \in \OO(2^{|\RR|})$). This
yields an $\OO(2^{\OO(|\RR|^2)})$ time complexity. Note that the
complexity of $\CC'$ is of the same order.

Finally, let us consider the construction of $\DD$. The number of
states of $\DD$ is in $\OO(2^{2^{|\RR|}})$ and the number of transition
rules in $\OO(2^{2^{\OO(|\RR|^2)}})$. The time to build $\DD$ is the
time to build $\CC'$ plus the time to compute the rules of $\DD$. The
former is neglectable with respect to the latter, which can be done in 
\[
\OO(2^{2^{\OO(|\RR|^2)}} \cdot |\CC'|) = \OO(2^{2^{\OO(|\RR|^2)}})
\]
time.

As emptiness can be decided in polynomial time with respect to the size
of the automaton, we conclude with the following theorem.

\begin{theorem}
\label{complexity-NF}
It can be decided in double exponential time whether a left-linear TRS
belongs to $\CBNNFg$.
\qed
\end{theorem}

Although the saturation process for $\RRs$ and $\RRnv$ is much simpler,
our constructions do not give better complexity results for deciding
membership in $\CBNNFs$ or $\CBNNFnv$. Nevertheless, for $\CBNNFs$
(which coincides with the class of strongly sequential TRSs, see
\cite{DM97}) a lower complexity bound is known. Comon~\cite{C95,C99} 
showed that it can be decided in exponential time whether a left-linear
TRS is strongly sequential. He uses an automaton for $\omega$-reduction
which plays the same role as our $\CC_{\NF_{\RR_\bullet}}$ automaton.
Since there is no satisfactory notion of $\omega$-reduction
corresponding to the approximations mappings $\mathrm{nv}$ and
$\mathrm{g}$, it remains to be seen whether the result of
Theorem~\ref{complexity-NF} can be improved.

\section{Call by Need Computations to Root-Stable Form}
\label{CBNRS}

In this section we assume that $\RR$ and $\SS$ are linear growing
TRSs over signature $\FF$ and
$\GG = \FF \cup \{ f^\circ \mid f \in \FF \}$. Our first goal is to
construct a tree automaton that recognizes the set $\RS_{\SS^\circ}$ of
root-stable ground terms with respect to the TRS
$\SS^\circ = \SS \cup \{ l^\circ \to r \mid l \to r \in \SS \}$.
Consider the automaton $\BB(\SS^\circ)$ defined in Section~\ref{step1}.
To this automaton we add a single final state $q_f$ and transition
rules $f(\state{l_1}, \dots, \state{l_n}) \to q_f$ and
$f^\circ(\state{l_1}, \dots, \state{l_n}) \to q_f$ for every left-hand
side $f(\seq{l})$ of a rewrite rule in $\SS$. One easily verifies that
the resulting automaton, which we denote by $\AA_{\REDEX_{\SS^\circ}}$,
accepts the set of ground redexes of $\SS^\circ$. Applying the
construction in Section~\ref{step3} to $\AA_{\REDEX_{\SS^\circ}}$ and
$\BB(\SS^\circ)$ results in an automaton
$\CC_{\REDEX_{\SS^\circ}}(\SS^\circ)$ that accepts all ground terms in
$\TT(\GG)$ that rewrite in $\SS^\circ$ to a term in
$\REDEX_{\SS^\circ}$, in other words, all non-root-stable terms of
$\SS^\circ$. From this we obtain the desired tree automaton
$\AA_{\RS_{\SS^\circ}}$ by a subset construction. Formally, the states
of $\AA_{\RS_{\SS^\circ}}$ are subsets of states of
$\CC_{\REDEX_{\SS^\circ}}(\SS^\circ)$, the final states are those
subsets that do not contain $q_f$ (the unique final state of
$\CC_{\REDEX_{\SS^\circ}}(\SS^\circ)$) and the transition rules of
$\AA_{\RS_{\SS^\circ}}$ are defined as expected.

Next we apply the constructions in Section~\ref{step3} to
$\AA_{\RS_{\SS^\circ}}$ and $\BB(\RR)$. This yields an automaton
$\CC_{\RS_{\SS^\circ}}(\RR)$ that accepts all ground terms that rewrite
in $\RR$ to a term in $\RS_{\SS^\circ}$. The construction in
Section~\ref{CBNNF} needs some modifications. We obtain
$\CC'_{\RS_{\SS^\circ}}(\RR)$ from $\CC_{\RS_{\SS^\circ}}(\RR)$ by
adding a fresh copy of $\BB(\RR)$ (we don't need the state $q_r$ here)
as well as the automaton $\CC_{\REDEX_{\SS}}(\SS)$. Let $Q_f'$ be the
set of final states of $\CC_{\REDEX_{\SS}}(\SS)$.
Concerning the construction of $\DD$, instead of
$P^2 = \{ \state{x} \}$ (in the case that there exists a rewrite rule
in $\RR$ with left-hand side $f(\seq{l})$ such that $\new{l_i} \in S_i$
for all $i \in \{ 1, \dots, n \}$) we define $P^2$ as a non-empty
subset of $f^\circ(S_1, \dots, S_n){\downarrow}$. Let us denote the
resulting automaton by $\DD'(\RR,\SS)$. The final states of
$\DD'(\RR,\SS)$ are those pairs $[S,P]$ that satisfy
$S \cap Q_f' \neq \varnothing$ and $P \subseteq Q_f$. The former
condition ensures that only non-$\RR$-root-stable terms are accepted.
The proofs of the following statements are easy modifications of the
corresponding ones in Section~\ref{CBNNF}. (In the proof of
Lemma~\ref{final-P2'} we take $P^2 = P \cap s^\circ{\downarrow}$ in the
case that $s$ is a redex.)

\begin{lemma}
\label{final-P1'}
Let $s \in \TT(\FF)$. If $s \to_{\Gamma_{\DD'}}^* [S,P]$ then for every
redex position $p$ in $s$ there exists a state $q \in P$ such that
$q \in s[(s_{|p})^\circ]_p{\downarrow}$.
\qed
\end{lemma}

\begin{lemma}
\label{final-P2'}
Let $s \in \TT(\FF)$. If
\[
P \subseteq \bigcup_{p \in \RR(s)} s[(s_{|p})^\circ]_p{\downarrow}
\]
such that $P \cap s[(s_{|p})^\circ]_p{\downarrow} \neq \varnothing$ for
all $p \in \RR(s)$ then $s \to_{\Gamma_{\DD'}}^* [S,P]$ where
$S = s{\downarrow}$.
\qed
\end{lemma}

\begin{lemma}
\label{final-P3'}
Let $s \in \TT(\FF)$. If $s \to_{\Gamma_{\DD'}}^* [S,P]$ and
$P \subseteq Q_f$ then, for every redex position $p$ in $s$,
$s[(s_{|p})^\circ]_p \in (\to_\RR^+)[\RS_{\SS^\circ}]$.
\qed
\end{lemma}

\begin{lemma}
\label{final-P4'}
Let $s \in \TT(\FF)$. If
$s[(s_{|p})^\circ]_p \in (\to_\RR^+)[\RS_{\SS^\circ}]$ for every redex
position $p$ in $s$ then there exists a set $P \subseteq Q_f$ such
that $s \to_{\Gamma_{\DD'}}^* [S,P]$ with $S = s{\downarrow}$.
\qed
\end{lemma}

\begin{corollary}
Let $s \in \TT(\FF)$. We have $s \in L(\DD'(\RR,\SS))$ if and only if
$s$ is non-$\SS$-root-stable and
$s[(s_{|p})^\circ]_p \in (\to_\RR^+)[\RS_{\SS^\circ}]$ for every redex
position $p$ in $s$.
\qed
\end{corollary}

\begin{theorem}
Let $\RR$ be a left-linear TRS and
$\alpha, \beta \in \{ \mathrm{s}, \mathrm{nv}, \mathrm{g} \}$. We have
$\RR \in \CBNRS_{\alpha,\beta}$ if and only if
$L(\DD'(\RR_\alpha,\RR_\beta)) = \varnothing$.
\end{theorem}
\begin{proof}
Note that $\RR_\alpha$ and $\RR_\beta$ are linear growing TRSs. We have
$\RR \notin \CBNRS_{\alpha,\beta}$ if and only if there exists a
non-$\RR_\beta$-root-stable term $s$ in $\TT(\FF)$ without
$(\RR_\alpha,\RR_\beta)$-root-needed redexes. The latter is equivalent
to $s[(s_{|p})^\circ]_p \in (\to_{\RR_\alpha}^+)[\RS_{\RR_\beta^\circ}]$
for every redex position $p$ in $s$. According to the preceding
corollary this is equivalent to $s \in L(\DD'(\RR_\alpha,\RR_\beta))$.
We conclude that $\RR \in \CBNRS_{\alpha,\beta}$ if and only if
$L(\DD'(\RR_\alpha,\RR_\beta)) = \varnothing$.
\qed
\end{proof}

The following complexity result is obtained by a similar analysis to
the one in Sect.~\ref{complexity}. Note that the nested saturation
process does not give rise to an extra exponential since saturation
increases only the time but not the space complexity by an exponential.

\begin{theorem}
\label{complexity-RS}
It can be decided in double exponential time whether a left-linear TRS
belongs to $\CBNRS_{\alpha,\beta}$ for all
$\alpha, \beta \in \{ \mathrm{s}, \mathrm{nv}, \mathrm{g} \}$.
\qed
\end{theorem}

\bibliography{references}

\end{document}